# A-site Ordering versus Electronic Inhomogeneity in CMR-Manganite Films

V. Moshnyaga[1*], L. Sudheendra[1], O. I. Lebedev[2], K. Gehrke, O. Shapoval[3], A. Belenchuk[3], S.A. Köster[1], B. Damaschke[1], G. van Tendeloo[2], K. Samwer[1]

[1] *Erstes Physikalisches Institut, Universität Göttingen, Friedrich-Hund-Platz 1, D-37077 Göttingen, Germany;*
[2] *EMAT, University of Antwerp (RUCA), Groenenborgerlaan 171, B-2020 Antwerpen, Belgium;*
[3] *Institute of Applied Physics, Academiei Str. 5, MD-2028, Chisinau, Moldova;*

Epitaxial La$_{3/4}$Ca$_{1/4}$MnO$_3$/MgO(100) (LCMO) thin films show unusual rhombohedral (R-3c) structure with a new perovskite superstructure due to unique ordering of La and Ca at the A-site positions. Very sharp insulator-metal and para-ferromagnetic phase transitions at temperatures up to $T_{MI} \sim T_C = 295$ K were observed. The ordered films were electronically homogeneous down to 1 nm scale as revealed by scanning tunnelling microscopy/spectroscopy. In contrast, orthorhombic and A-site disordered LCMO demonstrate broadened phase transitions as well as mesoscopic phase separation for $T \ll T_C$. The unique La/Ca ordering suppresses cation mismatch stress within one super-cell, a~1.55 nm, enhancing electronic homogeneity. Phase separation scenario seems not to be a unique mechanism for CMR as very large CMR=500 % was also observed in A-site ordered films.

PACS numbers: 75.47.Lx, 68.37.-d, 71.30+h

It is commonly believed that the basic physics of colossal magnetoresistant[1] (CMR) perovskite manganites, Ln$_{1-x}$Ae$_x$MnO$_3$ (Ln$^{3+}$=La, Nd, Sm, Pr etc. and Ae$^{2+}$=Ca, Sr, Ba), is governed by the interplay between spin-, charge- and lattice degrees of freedom. The resulting phase diagram contains a number of competing electronic and structural phases[2-4]. The Curie temperature, $T_C$, and metal-insulator transition temperature, $T_{MI}$, is controlled by A-site substitutions in the perovskite lattice with corresponding changing of the average ionic radius of the A-site cation, $<r_A>$, and the tolerance factor, $t = (r_A + r_O)/\sqrt{2}(r_B + r_O)$ [2-4]. Two types of A-site occupancy are reported for perovskite manganites. First, random distribution of A-site cations like in a solid solution is shown for CMR manganites[5], like LCMO. The disorder not only reduces $T_C$[5] due to size mismatch strain fields and corresponding Jahn-Teller (JT) distortions of the MnO$_6$ octahedra[6] but also causes electronic inhomogeneity or phase separation (PS)[7-10]. PS has been modelled both for cation disorder[11] and stress[12]. CMR effect was attributed to the disorder induced PS, imposed on the first order phase transition[13]. Secondly, the A-site ordered configuration was reported for "half-doped" manganites, Ln$_{0.5}$Ba$_{0.5}$MnO$_3$[14-16]. The LnO and BaO layers, alternatively stacked along the c-axis, provide the absence of random Coulomb potential and/or local strain in the MnO$_2$ sheets. Such an ordering stabilizes the long-range charge/orbital ordered (COO) insulating phase at high temperatures up to $T_{CO}$=500 K (for Ln=Y).

To the best of our knowledge there are no reports on the A-site ordered and optimally doped (0.2<x<0.4) CMR manganites. Prototypic middle bandwidth manganite as $La_{0.7}Ca_{0.3}MnO_3$ with relatively small average radius of A-site cation, $<r_A>$=1.205 Å, and tolerance factor, t=0.97<1, possesses orthorhombic ($P_{nma}$) structure at room temperature and transforms to a rhombohedral (R-3c) structure at 700 K[6]. In this letter we report LCMO films with pseudo-cubic (R-3c) structure and new type of perovskite superstructure due to La/Ca ordering. We show that a unique ordering at the A-site positions not only brings about long range Coulomb order, but also suppresses cation mismatch strain within the superstructure unit cell.

LCMO films with nominal Ca-doping level x=0.25 were epitaxially grown on MgO(100) substrates by a metalorganic aerosol deposition (MAD) technique[17,18]. The thickness of the films were d=80-90 nm as determined from small angle X-ray scattering (XRD). XRD analysis shows "cube-on-cube" epitaxy of LCMO on MgO(100) and indicates a stress-free state of the film: the film lattice parameter, $a_p$=0.3875-0.3872 nm, is very close to bulk value[19]. Remarkably, in LCMO films a pseudo-cubic or rhombohedral (*R-3c*) crystalline structure was observed by transmission electron microscopy (TEM) and electron diffraction (ED) analysis, shown in Fig. 1. The brighter basic reflections in both patterns show a square arrangement with a ≈ $a_p$. Weak satellite reflections on both ED patterns suggest a modulated structure with a modulation vector $q_1$ ≈ 1/4 $[011]*_C$ in the cross-section (Fig. 1c) and $q_2$ ≈ 1/2 ($[001]*_C$ in plain-view ED pattern (Fig. 1a). High resolution (HREM) plain-view and cross-section images are shown in Fig. 1b & 1d, respectively. In the plain view a twinned structure was seen; therefore the modulation in the ED patterns is only one- dimensional, but appears in two symmetry related directions. The La(Ca)O layers (indicated by arrows in Fig. 1b & 1d are imaged as rows of brighter dots, whereas LaO layers are imaged as rows of less brightness dots in Fig. 1b. The cross-section HREM image in Fig. 1d gives even more evidence for this modulated structure. The Fourier transform (not shown here) of this modulated area exhibits a pattern very similar to the corresponding ED pattern in Fig. 1c. The modulation shows up as alternative blocks of *$2a_p$* x *$a_p$* size of a different brightness, indicating an ordering of La and Ca-columns.

In Fig. 2 a model for the La/Ca ordering is shown, which is in agreement with the measured ED and HREM. The La-Ca ordering involves an alternating stacking of La-O and La(Ca)-O layers along the $[010]_c$ direction. The La and Ca are ordered within the La(Ca)-O layer along the $[100]_c$ direction: two La atoms alternate with two Ca atoms (see $[001]_c$ view in Fig. 2b). The layering of LaO and (La,Ca)O planes becomes evident along the $[011]_c$ direction. In this direction the film can be viewed as composed of two repeating La-Mn-O (LMO) layers followed by the two layers of La-Mn-Ca-Mn-O (LMCMO). Both layers together form the body-centred cubic superstructure unit cell. The theoretical cation stoichiometry, $La_{3/4}Ca_{1/4}Mn_1O_3$, agrees with nominal composition. The simulated ED patterns in Fig. 2 correlate very well with those measured (Fig. 1). The image simulations (the insets in Fig. 2a, and 2b), based on the proposed model, are also in agreement with the experimental images.

We also checked for PS in cation ordered LCMO (O-film) in comparison with disordered a orthorhombic LCMO (D-film)[20] using scanning tunnelling microscopy/spectroscopy (STM/STS)[9]. The surface of the O-film (Fig. 3a) looks very smooth and flat with average surface roughness, RMS=0.2 nm. One-unit cell height steps are clearly seen in between 100-200 nm wide terraces. The D-film (Fig. 3b) shows crystalline grains with diameter ~50 nm and RMS=2 nm. In Fig. 3c, & 3d we present the distribution of the local tunnelling conductance, $\sigma=(dI/dV)_{V=0}$, measured in the ferromagnetic state for T=115 K<<$T_C$ for the O- and D-films, respectively. The O-film (Fig. 3c) shows spatially homogenous, at least down to 1 nm scale (the limit of our STS setup), tunnelling conductance of metallic type i.e. $\sigma \neq 0$. The conductance map of the D-film (Fig. 3d) is inhomogeneous – insulating regions, $\sigma \approx 0$, of about 5-20 nm size ("black clouds" in Fig. 3d) are intermixed with metallic (red) regions. Thus, even for T<<$T_C$ the D-film shows a mesoscopic PS, whereas the O-film does not. In Fig. 4a we compared temperature dependence of the resistivity, $\rho(T)$, for the O-film with $T_{MI}$=272 K and $T_C$=264 K and that for a bulk orthorhombic LCMO[3] with the same doping level x=0.25. The metal-insulator transition, characterized by the temperature coefficient of the resistance $TCR_{max}=100\%*(1/R)(dR/dT)$[21], is very sharp for the cation ordered LCMO film, yielding $TCR_{max}$=37 %/K. In contrast the bulk orthorhombic LCMO possesses broadened MI- transition, $TCR_{max}$=12 %, and significantly lower $T_{MI}$~240 K.

We have observed a new perovskite superstructure due to A-site cation ordering in LCMO films. A more symmetric pseudo-cubic (R-3c) crystalline structure builds up with enlarged super-lattice cell in [001] direction i.e. $a_s=4a_p$=1.55 nm. We analysed the projections of perovskite unit cells from the cross-section HREM pictures by assuming the positions for La- (shown as blue spheres), Ca- (red) and Mn- (white) atoms and superpose them on the cross-section HREM in Fig. 2c. Remarkably, the local stresses due to La/Ca cation size mismatch, which are crucial for PS and magnetotransport phenomena, are compensated within one super-cell. Basic LMO and CMO cells, approximated as quadrangles around La (B-cell) and Ca (R-cell) atoms, respectively, are evidently not equal: their lengths relate exactly to their lattice parameters, i.e. $a_{LMO}$=0.390 nm and $a_{CMO}$=0.373 nm. This indicates a stress-free state of R- and B-cells. Along [001]- and [010]-axes basic cells are linked by the intermediate, B1-cells, which accommodate LMO/CMO local stresses via axial deformation with intervening tensile, $B1_+$, and compressive, $B1_-$, strains of (Mn-Mn)-distances, probably due to the change of (Mn-O-Mn)-angle. The sequence of strained domains is the following:

$$[R-(B1_+)-B-(B1_-)]-[(R-(B1_+)-B-(B1_-)]-\ldots \quad (1)$$

Along the diagonals of super-cell, the local stress has rather biaxial nature resulting in parallelogram-like deformation, $(B1_{+-})$, with the following sequence of strain domains:

$$[R-(B1_{+-})-R-(B1_{+-})]-[(R-(B1_{+-})-R-(B1_{+-})]-\ldots \quad (2)$$

The resulting distribution of local stresses within one super-cell is shown in Fig. 2d. One can see tensile-like (blue) and the compressive-like (red) stressed perovskite cells arranged in a "zigzag-stripe" fashion along the [011] diagonal of the super-cell. The overall stress, being the sum of red and blue regions, is fully compensated within one super-cell. In another words, the ordered LCMO film along the [011]-direction looks as infinite superlattice, [2"LMO"/2"$La_{0.5}Ca_{0.5}MnO_3$"]$_N$, in which small lattice misfit $\varepsilon=(a_{LMO}-a_{LCMO})/a_{LMO}=1.5$ % can be accommodated by the above shown mechanism. Such a layering looks energetically more favourable than a simple [LMO/CMO]$_N$ layering, which would result in a significantly larger lattice misfit, $\varepsilon=4.5$ %.

We believe that the revealed very short scale stress accommodation mechanism drives the formation of a cation-ordered perovskite superstructure with enlarged super-cell, $a_s=4a_p$. This modifies the crystalline structure of the film from the bulk orthorhombic to a more symmetric R-3c structure, which diminishes lattice distortions. Moreover, it provides a natural link between the disorder[11] and stress[12] in computational models for the phase separation problem (Fig. 3). Short scale stress relaxation within (R-3c) structure in the A-site ordered LCMO films yields very sharp phase transition and homogeneous ferromagnetic (FM) metallic ground state for $T<T_C$. In contrast, cation disorder within orthorhombic LCMO exerts stress on a much larger, presumably, mesoscopic scale. This results in a structurally correlated nano-scale regions with cooperative Jahn-Teller (JT) effect[22,23] and localized charge carriers in accordance with observed broadening of phase transitions and metallic and insulating domains for $T<T_C$ in Fig. 3d. Interestingly, very large CMR=100%[(R(0)-R(5T)]/R(5T)$\approx$500 %, observed for A-site ordered LCMO (see Fig. 4), means that mescoscopic phase separation[11] seems to be not necessary attribute for the CMR effect at the FM/paramagnetic transition. This appears to be in accordance with "CMR1" effect[24], which occurs in the absence of quenched disorder close to antiferromagnetic(AF)/charge ordering phase boundary as field induced AF/FM transition.

Our experimental results allow to formulate a common "superlattice" rule to search for the A-site ordered manganite films:

$$[(mLn\text{-}Mn\text{-}O)/(2(Ln\text{-}Mn\text{-}B\text{-}Mn\text{-}O)]_N \qquad (4)$$

with m=0,1,2,3,..; Ln=Y, La, Nd, Pr…; and B=Ca, Sr, Ba. The ordered variants of LCMO films for different numbers of LMO layers "m" can be obtained as following: for m=1 we get $La_{2/3}Ca_{1/3}MnO_3$; for m=2 – $La_{3/4}Ca_{1/4}MnO_3$; m=3 – $La_{4/5}Ca_{1/5}MnO_3$; m=4 – $La_{5/6}Ca_{1/6}MnO_3$. Such a growth strategy provides a tool to check the relevance of earlier estimations[5] that the disorder-free manganite with large average cation radius (Ba) would possess very high $T_C$=530 K. Interestingly, the layer-by-layer MAD-grown $La_{1-x}Ca_xMnO_3$ films show an increase of the $T_{MI}$ of about 40 K (shown in Fig. 4b) as compared to the bulk samples[3], reaching room temperature, $T_{MI}$=295 K. There are few earlier reports[25,26] on the possibility to increase the transition temperature for La-Ca-Mn-O films, prepared by pulsed laser ablation, up to room temperature region. However, there was no evidence on the origin of the observed effect. Our results indicate that the increase of the transition temperature could be achieved by controlling the structure and cation ordering in optimally doped manganite films, showing CMR effect.

The authors thank Ch. Renner for the help in STM/STS and E. Dagotto for valuable comments. Deutsche Forschungsgemeinschaft via SFB 602, TPA2 is acknowledged. L.S. thanks the A. von Humboldt Foundation for the fellowship.

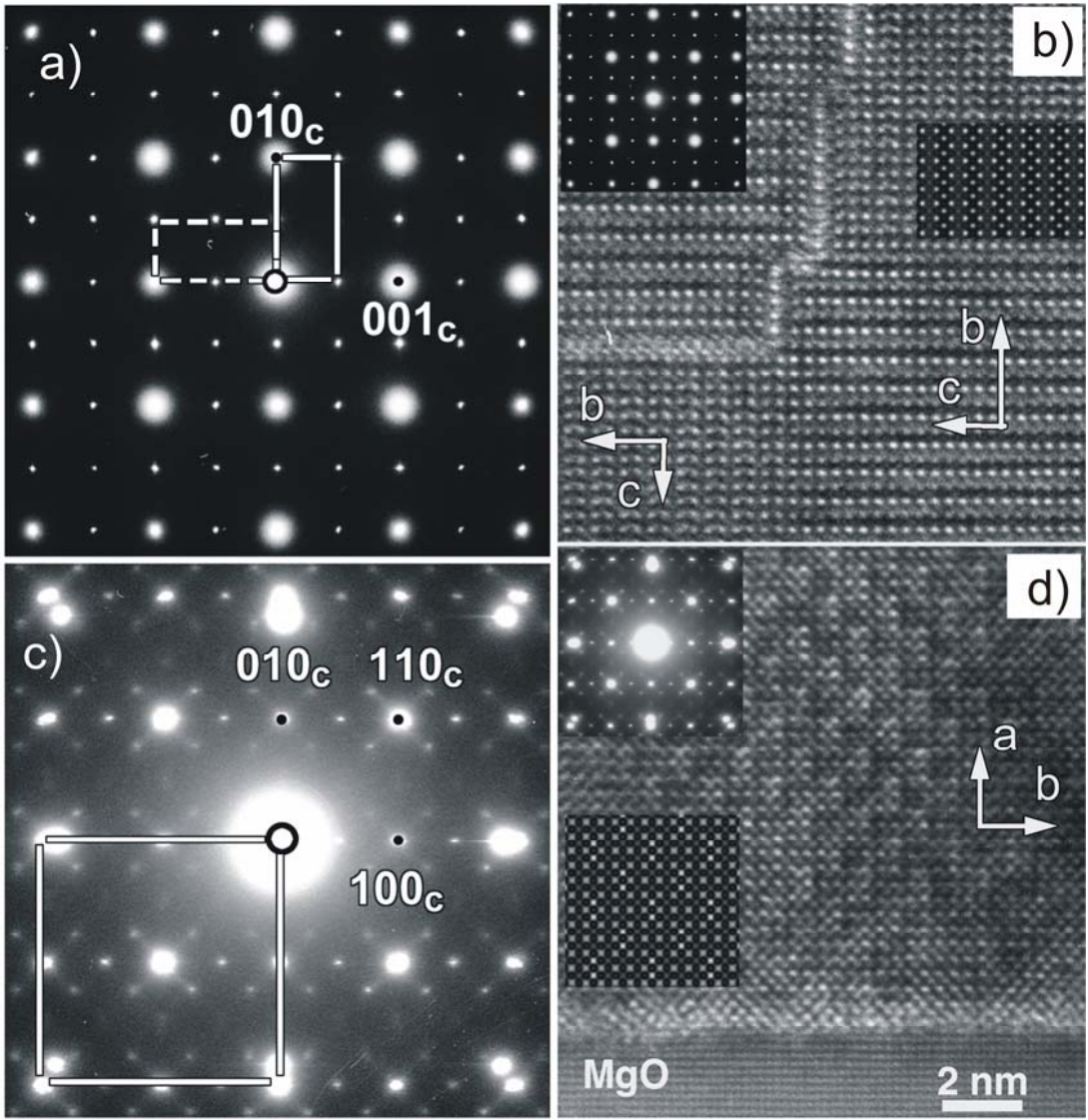

FIG. 1 Plan-view (a) and cross-section (c) electron diffraction (ED) patterns of the cation ordered La$_{0.75}$Ca$_{0.25}$MnO$_3$ film with modulated structure. The ED pattern obtained from the MgO substrate indicated by white square in (c). The white rectangles in (a) correspond to the twinned LCMO structure. Plan-view (b) and cross-section (d) HREM images of the same film. The Fourier transforms from single domain of the modulated area given as insert and exhibit a pattern similar to the corresponding ED patterns in Fig. 1a) and 1c).

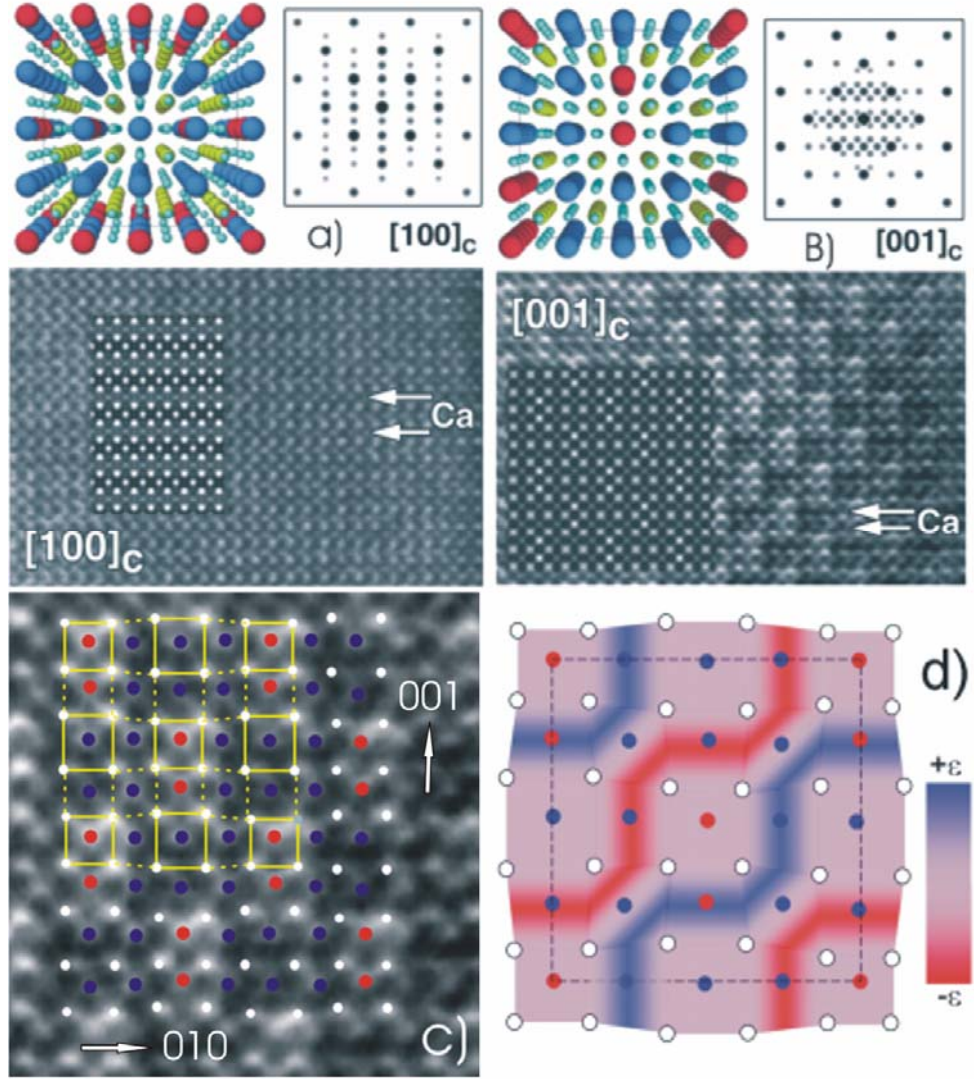

FIG. 2 Structural model of the La-Ca ordered $La_{0.75}Ca_{0.25}MnO_3$ phase, taking into account experimental evidence: (a) - represents the projection along $[100]_c$ direction and (b) represents the projection along $[001]_c$ direction. The corresponding calculated ED patterns and simulated HREM images obtained at a defocus value $\Delta f = -30$ nm and thickness $t = 18$ nm based on the proposed model (given as inset in HREM images) show a good agreement with an experimental ones. In (c) a fragment of HREM image (b) is shown with superimposed single perovskite cells, containing A-cations (red - Ca and blue - La) and Mn (white). The distribution of local stress is shown in (d) with compressive-like (red stripes) and tensile-like (blue stripes). The overall stress is compensated within one super-cell, $a_S=4a_p=1.55$ nm

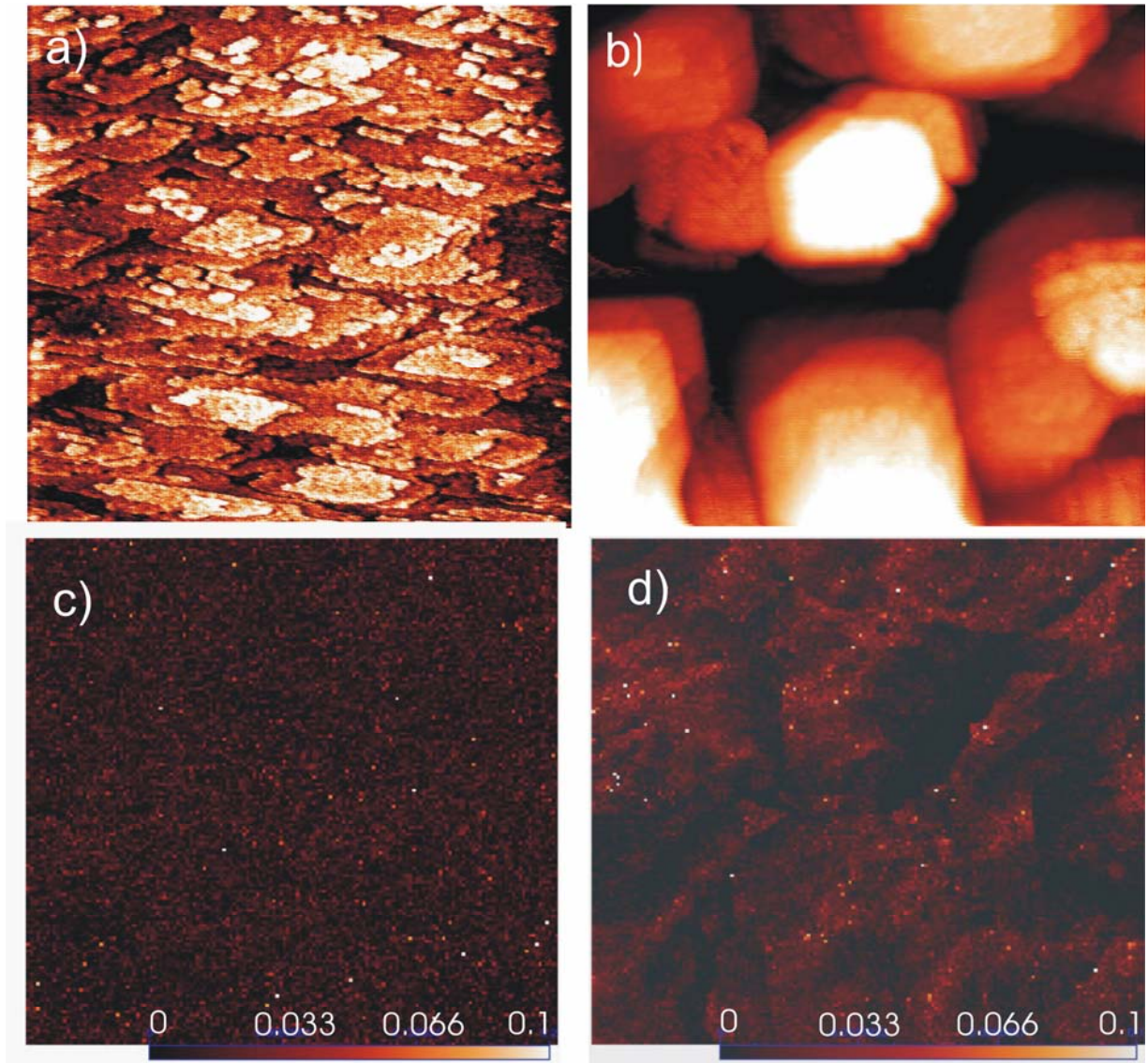

FIG. 3 Scanning tunneling microscopy (STM) morphology image (500x500 nm$^2$) of a pseudo-cubic LCMO ("O-film") (a) and 250x250 nm$^2$ STM morphology scan of the orthorhombic "D-film" (b). The tunneling conductivity map of "O-film" (c)- and "D-films" (d), obtained from the same regions as in Fig. 1(a) and (b), respectively. Electronically homogeneous behaviour in "O-film" and metal-like (red) and insulator-like (black) coexistence in "D-film" are evident. Scale represents the conductance in nA/V.

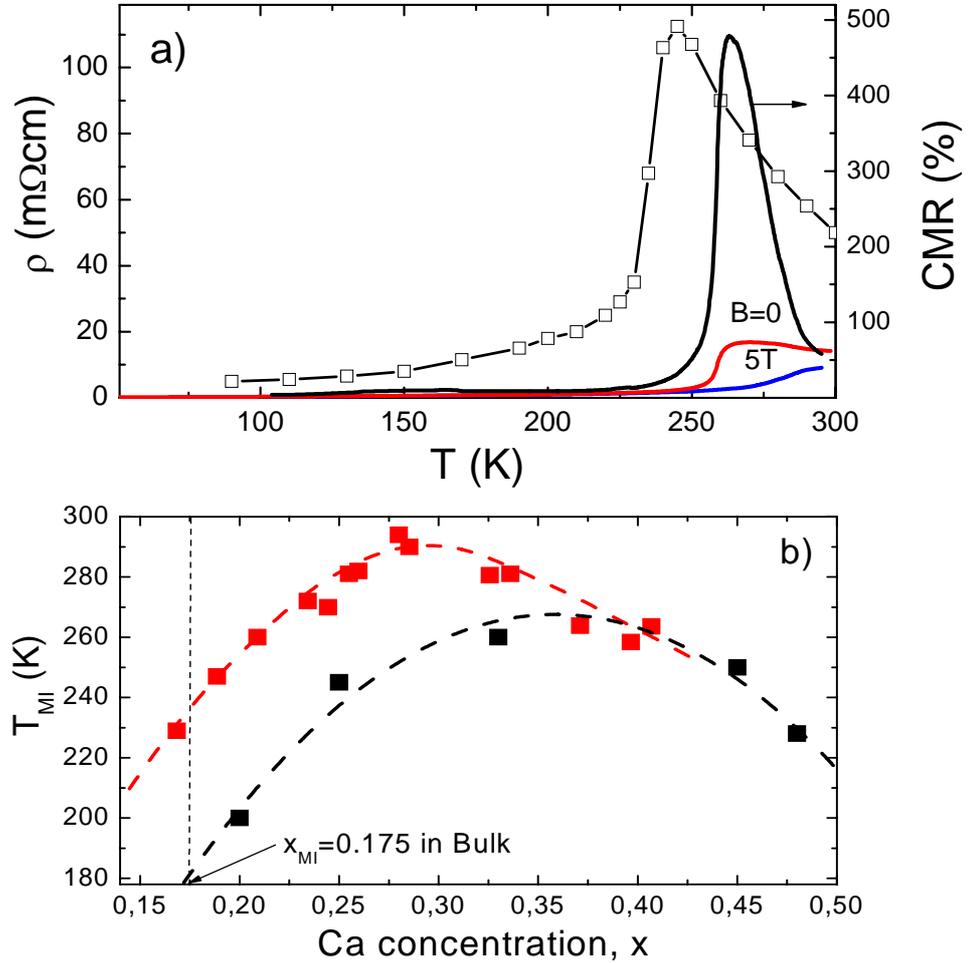

FIG. 4. (a) - Temperature dependence of the resistivity, $\rho(T)$, for cation ordered film (red curve by B=0 and the blue one for B=5 T) and for a bulk sample (redrawn from Ref. 3, open quadrates) of the same $La_{0.75}Ca_{0.25}MnO_3$ composition. The right scale relates to the evaluated temperature dependence of CMR=100%(R(0)-R(5T))/R(5T); (b) – Metal-insulator transition temperatures, $T_{MI}$, for MAD layer-by-layer grown LCMO/MgO films (red quadrates) for bulk samples (redrawn from Ref. 3, black quadrates) as a function of Ca-doping level. The connecting lines are drawn as guide to the eye.